# DYNAMIC SMART CHARGING THROUGH INTEGRATED AGENT-BASED SIMULATION AND OPTIMIZATION MODELING: IMPACTS ON FLEET OPERATIONS, EMISSIONS, AND POWER GRIDS

**Lin Su**
Maseeh Department of Civil, Architectural and Environmental Engineering
The University of Texas at Austin
301 E. Dean Keeton St, Stop C1761, Austin, TX, 78712
sulin@utexas.edu
ORCID: https://orcid.org/0000-0001-5507-0389

**Kara M. Kockelman, Ph.D., P.E.**
(Corresponding Author)
Dewitt Greer Professor in Engineering
Maseeh Department of Civil, Architectural and Environmental Engineering
The University of Texas at Austin
301 E. Dean Keeton St, Stop C1761, Austin, TX, 78712
kkockelm@mail.utexas.edu
Tel: 512-471-0210

## ABSTRACT

Coordinated electric vehicle (EV) charging can improve the alignment between transportation operations and power-system conditions as fleet and personal vehicle electrification expands. This study develops an integrated simulation-optimization framework that embeds demand-aware charging (DAC) and grid-responsive charging within an agent-based model, enabling dynamic feedback between optimized vehicle (dis)charging decisions and evolving transportation-system states. DAC schedules shared autonomous electric vehicle (SAEV) charging based on anticipated fleet use and practical charging availability, while grid-responsive strategies coordinate unidirectional smart charging (V1G) and bidirectional vehicle-to-grid (V2G) operations with time-varying grid conditions. Using the Austin 6-county region, the framework evaluates a 15,000-vehicle SAEV fleet under DAC, V1G, fleet-only V2G (Fleet-V2G), and combined fleet-plus-household V2G (Combined-V2G). Results show that DAC concentrates 35% of daily charging between 10 AM and 2 PM, with no charging occurring before 6 AM. Grid-responsive strategies allocate 29% to 42% of charging to midday hours, when electricity prices and grid stress are generally lower. Fleet-V2G provides selective evening fleet discharge, while Combined-V2G extends grid support through household EV participation. All coordinated strategies maintain broadly comparable ride-hailing service performance. Under favorable pricing structures, V2G reduces net $CO_2$ emissions from 45 kg/SAEV/day under DAC to below 40 kg and lowers emission intensity by up to 14% to 0.14 kg/VMT. Passenger-service operations remain the primary driver of fleet profitability, while smart-charging savings provide modest gains. DAC and V1G offer practical near-term options, whereas high upfront V2G costs may hinder widespread adoption. The realized benefits depend strongly on tariff design, vehicle availability, and V2G infrastructure deployment.

## 1. BACKGROUND

Vehicle electrification has emerged as a worldwide trend, driven by advances in battery technology, declining vehicle costs, and supportive policies, including U.S. state-level programs such as New York's Drive Clean Rebate, Connecticut's CHEAPER program, and Massachusetts's MOR-EV program (Kennedy et al., 2025). Global electric vehicle (EV) sales reached a record high of over 20 million vehicles in 2025, accounting for 25% of all new car sales worldwide (IEA, 2026). In addition, a 300-km-range BEV is estimated to produce approximately 54% lower lifetime $CO_2$-eq emissions than a comparable internal combustion engine (ICE) vehicle and 32% lower emissions than an equivalent PHEV (IEA, 2024). Electrification is also occurring across commercial vehicles. For example, Lyft aims for 100% EVs on its platform by 2030, while Uber targets 100% zero-emission rides globally by 2040, with a 2030 goal for U.S., Canadian, and European cities (Lyft, 2020; Uber, 2020). Meanwhile, emerging technologies, including vehicle automation, shared mobility platforms, and electric powertrain development, are reinforcing this trend. Waymo has deployed around 2,500 fully autonomous EVs across San Francisco, Los Angeles, Phoenix, and Austin as of late 2025, and is expected to reach 3,500 vehicles by 2026 (Campbell, 2025a; Waymo, 2025).

The rapid growth of personal and fleet electrification places rising pressure on charging infrastructure and power systems, requiring both charging infrastructure expansion and strategic deployment approaches, such as public-private collaboration (Su et al., 2024). In 2023, researchers at the National Laboratory of the Rockies estimated that about 1 million public Level 2 charging ports and 182,000 public DC fast charging (DCFC) ports would be needed in the U.S. by 2030 to support accelerated EV adoption (Wood et al., 2023). Without coordinated management, uneven spatial and temporal charging demand can exacerbate regional grid stress and local congestion at charging hubs. Bernard et al. (2025) found that public charging demand is highly responsive to price signals, enabling charging to shift away from peak periods toward lower-cost hours (e.g., midday, thanks to solar panels). As utilities and planners seek scalable solutions to maintain system reliability while supporting increasing charging demand, coordinated charging and grid-responsive smart charging have emerged as promising strategies for mitigating peak loads, improving charge scheduling, and supporting grid operations.

EV charging strategies differ in both the information they require and the level of control they provide. Conventional uncontrolled charging assigns a vehicle for charging primarily when its state of charge (SoC) falls below a specified threshold, without considering time-varying travel demand or the opportunity cost of removing a vehicle from passenger service. Thus, charging may occur during high-demand periods, reducing fleet availability and potential service revenue.

Smart-charging management (SCM) is the coordinated and dynamic control of EV charging in response to grid conditions, fleet operating needs, and electricity prices, potentially reducing costly distribution upgrades and EVCS deployment delays caused by limited grid capacity (U.S. DOE, 2025a). Smart charging strategies include managed unidirectional charging (V1G) and bidirectional vehicle-to-grid (V2G) service. While V1G primarily smooths grid loads, V2G offers greater flexibility by enabling EVs to supply stored energy back to the grid during system stress or high-price hours. NovaCharge (2024) similarly notes that fast charging imposes concentrated loads on the grid, whereas scheduling optimization can reduce both operating costs and alleviate grid stress. Large-scale participation in smart charging, especially V2G, spans household (HH) EVs and commercial fleets such as shared automated EVs (SAEVs).

Fleet EVs are well-suited for coordinated charging and V2G participation due to their predictable schedules, centralized management, continuous vehicle monitoring, and controllable operations. Driivz (2025) notes that coordinated V2G scheduling allows fleet operators to generate new revenue streams, helping offset electrification costs. Kumar (2024) further highlights that commercial fleets such as delivery vans, transit, or school buses are characterized by predictable schedules and long idle windows, making them promising providers of grid services. SAEV fleets represent a growing component of urban mobility and offer additional opportunities for energy flexibility. Campbell (2025) reported that Waymo's automated EV fleets reached a deadheading share of 44.3% in September 2025. Similarly, Raifman (2025) estimated that Waymo vehicles accumulated more than 304,000 unassigned passenger-service hours in March 2025. Although these idle periods reflect operational inefficiencies, they create substantial windows for smart charging.

While personal EVs offer greater long-term flexibility, their vehicle-grid integration (VGI) faces challenges due to unpredictable household travel, concerns about battery degradation, and the upfront cost of bidirectional chargers. Despite these barriers, their scales and long parking durations provide significant distributed energy storage potential, supported by National Household Travel Survey (NHTS) data, which indicate that household vehicles remain parked for over 95% of the daytime on average (U.S. DOT, 2024; Kane, 2024; Raifman, 2025). Empirical evidence further supports the feasibility and economic benefits of managed HH EV charging. Metcalfe et al. (2025) showed that managed charging can reduce peak household electricity demand by 42% while shifting nearly all charging to off-peak periods without increasing total electricity consumed, and generate consumer surplus, including annual bill savings ranging from £343 to £650 relative to a standard flat tariff. From the personal EV owner's perspective, lower charging costs, economic compensation, environmental benefits, and improved charging safety are primary incentives for V2G participation (Driivz, 2023; Zhang et al., 2024). At scale, these distributed resources can deliver substantial system-level value. The Electric Power Research Institute (EPRI) estimates that V2G could generate up to $1 billion in annual grid benefits in California by 2030 under high EV adoption scenarios of roughly 5 million EVs, half of which are V2G-enabled personal EVs (EPRI Journal, 2020). These findings highlight the need for modeling frameworks that can jointly represent household-level V2G availability and fleet-based charging and V2G operations to evaluate system-wide impacts.

### 1.1 Research Gaps and Challenges

Despite these promising findings across diverse EV groups, most existing studies focus on a single participating EV group, typically modeling either personal or household EVs (Ding et al., 2022; Zhang et al., 2024; Huang and Sandström, 2025; Nadimi and Goto, 2026) or commercial EV fleets (Liao et al., 2021; Dean et al., 2023; Aktar et al., 2023; Yu et al., 2024) in isolation. Few studies consider combined fleet-plus-household V2G (Combined-V2G) participation by both fleet and personal EVs, even though interactions between these groups can significantly influence grid performance, fleet operations, and energy markets. As a result, the system-wide costs, benefits, and operational trade-offs associated with coordinated participation across heterogeneous EV users remain underexplored.

Another gap concerns the operational baseline used to evaluate smart charging. Conventional uncontrolled charging represents a reactive SoC-based policy but overlooks the opportunity cost of charging fleet vehicles during high-demand periods. A demand-aware charging (DAC) strategy is expected to offer a more realistic operational baseline by coordinating charging around expected fleet use and practical charging availability without simplifying fleet charging decisions or relying on grid-side information.

Moreover, existing research often simplifies how EVs interact with both grid conditions and mobility needs. Many studies assume static traffic conditions or average travel patterns for EVs (Liao et al., 2021; Sun et al., 2021; Kobayashi et al., 2025), limiting their ability to capture operational constraints that shape real-world EV flexibility for V2G. Evaluations further indicate that behavioral non-adherence, time-varying traffic congestion, and grid uncertainty can reduce the realized value of optimized smart charging plans (U.S. DOE, 2025b; Brinkel et al., 2024; Jaworski et al., 2023). These limitations challenge examining smart charging strategies under dynamically evolving transportation environments, revealing a gap in existing VGI modeling approaches.

Addressing these gaps requires a modeling framework that integrates vehicle behavior with system-level optimization. A two-way simulation-optimization approach can support this need by allowing optimized (dis)charging schedules to inform vehicle behavior, while simulated travel patterns, network congestion, and evolving SoC distributions feed back into the (dis)charging optimization module. Agent-based models (ABMs) capture individual mobility decisions, interactions with the grid, and vehicle states, while optimization models identify cost-effective energy transactions under time-varying electricity prices and grid conditions. Their interaction captures deviations between planned and actual energy use, enables timely adjustments, and supports a more comprehensive evaluation of DAC, V1G, fleet-based V2G (Fleet-V2G), and fleet-plus-household V2G (Combined-V2G) strategies.

### 1.2 Contribution

This study addresses these gaps by evaluating coordinated smart charging for both personal EVs and a commercial SAEV fleet within a dynamic modeling environment. An integrated simulation-optimization framework is developed to jointly manage (dis)charging decisions for diverse EV groups. The framework advances existing VGI modeling approaches by embedding rolling 24-hour charging optimization modules within a large-scale agent-based simulation, enabling optimized energy decisions to continuously adapt to evolving travel demand and grid conditions.

This study also develops a DAC strategy that uses historical fleet-use patterns to schedule charging during lower-demand periods, ensuring fleet availability during peak hours. DAC can also incorporate practical constraints, such as personnel availability for connecting SAEVs to chargers. Thus, DAC remains a V0G strategy while providing a more operationally realistic baseline. This operational coordination makes DAC particularly relevant to SAEV fleets and provides a practical transition from reactive charging toward grid-responsive V1G and V2G.

The framework captures bidirectional feedback between optimized energy transactions and evolving travel behavior. Personal EVs participate in V2G services based on planned daily activities and SoC availability, while fleet EVs follow operational decisions for routing, service assignment, and vehicle repositioning. By modeling travel behavior, charging-infrastructure networks, electricity market conditions, and grid constraints, the framework enables a comprehensive assessment of economic outcomes, fleet performance, grid support potential, and pollutant-related impacts across multiple charging strategies.

Applied to a rapidly electrifying metropolitan region, the simulation-optimization model provides insights into how coordinated charging and V2G strategies influence fleet operations, grid-support behavior, and economic outcomes across alternative coordinated charging strategies. The results offer practical insights for fleet operators and planners seeking to design incentive programs, grid-management policies, and charging infrastructure deployment strategies.

The remainder of the paper is organized as follows. The next section presents the simulation-optimization methodology, including the agent-based travel simulation environment, DAC framework, personal EV discharge potential estimation, and the coordinated day-ahead smart charging workflow. The subsequent section describes the case study area and corresponding model configuration. The Results and Discussion section then compares outcomes across DAC, V1G, Fleet-V2G, and Combined-V2G. The paper concludes with key findings, policy-relevant insights, study limitations, and directions for future research.

## 2. METHODOLOGY

This study develops an integrated simulation-optimization framework that embeds two distinct optimization modules for DAC and smart charging within a large-scale ABM to evaluate alternative EV charging strategies. The simulation is implemented in POLARIS, a high-performance ABM that simulates agent activities, travel patterns, traffic conditions, and fleet operations across a 24-hour simulation day. DAC coordinates SAEV charging based on anticipated fleet use and practical charging availability, whereas grid-responsive V1G and V2G strategies are optimized based on time-varying electricity prices and grid conditions. The optimization module uses updated system states from POLARIS to determine charging and discharging schedules over a rolling 24-hour horizon.

The simulation and optimization components interact dynamically throughout the simulation. Updated travel and vehicle states from POLARIS inform subsequent charging and discharging decisions, while optimized energy schedules shape SAEV availability and fleet operations. POLARIS then captures the resulting travel responses, congestion dynamics, and deviations between predicted and realized vehicle energy use, which are incorporated into the next optimization cycle. This iterative structure allows charging and discharging decisions to adapt to evolving transportation-system conditions rather than relying on fixed daily schedules.

Within this framework, personal EVs have flexible V2G availability that aligns with the corresponding agent's daily schedules. Feasible HH EV discharge capacity is dynamically estimated based on vehicle availability, battery state, and owners' activity schedules, allowing V2G participation without disrupting planned travel. By adjusting the model's charging and discharging controls for fleet and personal EVs, the framework can simulate various smart charging strategies, including V1G, fleet-only V2G, and combined fleet-plus-household V2G, enabling comparison with the DAC scenario. These scenarios enable systematic comparison of reactive charging, fleet-side demand-aware coordination, and increasingly grid-responsive charging and discharging strategies.

The following subsections detail the agent-based simulation model, the DAC framework, the estimation of feasible discharge potential for personal EVs, and the day-ahead smart charging framework that integrates the CPLEX optimization module with ABM.

### 2.1 Agent-based Travel Simulation

The simulation environment for this work is POLARIS, an activity-based and agent-based modeling platform developed in C++ and designed for computationally intensive, large-scale transportation analysis (Auld et al., 2016). POLARIS has been widely applied in transportation simulation studies, including first-and-last-mile service analysis (Huang and Verbas, 2021), multi-stage fleet charging (Dean et al., 2023) and charging infrastructure planning (Dean et al., 2022; Su et al., 2025), and shared autonomous (electric) vehicle operations (Gurumurthy et al., 2021), highlighting its wide use across diverse transportation research.

POLARIS generates a synthetic population using demographic data from regional metropolitan planning organizations and the U.S. Census Bureau. The ADAPTS framework generates representative person-level characteristics, such as gender, age, and race, alongside household-level attributes including size, income, and number of employees (Auld and Mohammadian, 2009; Auld and Mohammadian, 2012). Each agent's daily activities are planned through activity generation models that consider activity types and demographic factors, and then organized using an activity-planning-order model to achieve temporal flexibility and planning horizons for different activities (Auld et al., 2011; Auld and Mohammadian, 2012). Trips and vehicles are loaded onto the POLARIS road network using a time-dependent traffic assignment model (Verbas et al., 2018). The model outputs detailed link-level trajectories for all trips within the study region and records traveler-and-vehicle-level travel information, such as charging activities, trip distance, and fleet ride requests (Moawad et al., 2021).

Within this framework, both fleet and personal EVs are modeled as mobility agents whose actions are driven by agents' needs. Fleet EVs follow operator-level strategies for trip assignment, repositioning, and maintenance. Ride requests are matched to available vehicles using a zone-based heuristic that checks detour distance, range feasibility, and service-quality constraints, such as maximum pickup delays (Gurumurthy et al., 2020). The model continuously updates each fleet EV's SoC and evaluates whether upcoming requests are feasible. A vehicle is assigned to new trips only when its SoC exceeds a minimum threshold and is sufficient for the predicted energy demand of the next ride (Gurumurthy et al., 2021). Personal EVs follow similar vehicle-level charging logic but are aligned with daily activity schedules and home-work activity patterns of their corresponding agents.

POLARIS enables a more realistic representation of agents' activities, mobility needs, and fleet operations. Key operational metrics, including vehicle-miles traveled (VMT), charging patterns, and regional configurations, are exported to the day-ahead optimization model in CPLEX. The simulation-optimization integration ensures that V2G participation remains operationally feasible, maintaining fleet service quality and respecting personal activity schedules while evaluating alternative smart charging strategies.

### 2.2 Demand-aware Fleet Charging

For high-use mobility fleets, charging decisions are closely coupled with passenger-service operations because vehicles undergoing charging are temporarily unavailable to serve ride requests. The operational opportunity cost of charging therefore varies throughout the day with rider demand. To capture this temporal tradeoff, this study develops a DAC strategy. This strategy coordinates fleet charging with time-varying fleet use while remaining independent of electricity prices and grid conditions. It therefore represents fleet-side operational coordination rather than grid-responsive smart charging, providing an intermediate strategy between conventional SoC-based charging and V1G or V2G control.

The DAC model uses fleet operational information generated by POLARIS. At each optimization step, the model determines fleet-level energy consumption over the planning horizon while accounting for expected fleet energy consumption, energy balance, fleet use, charger availability, and the target end-of-day SoC. The principal decision variable, $E_b^F(t)$, represents the aggregate energy purchased (in kWh) by the fleet at time step *t*. Table 1 summarizes the variables and assumptions used.

Table 1. Variables and Parameters for the Demand-Aware Charging

| Type | Symbol | Description |
|---|---|---|
| Set | $t$ | Charging time step, $t \in [0, T]$ |
| Decision Variables | $E_b^F(t)$ | Aggregate energy bought (charged) by the fleet at time step *t* (kWh) |

| Type | Symbol | Description |
|---|---|---|
| State Variables | $Q_F(t)$ | Aggregate energy stored in fleet's batteries at time step $t$ (kWh) |
| | $P_b^{Fmax}$ | Fleet's peak power demand when charging/buying energy (kW) |
| Parameters | $E_b^{Fmax}(t)$ | Maximum energy the fleet can buy at time step $t$ (kWh) |
| | $\widetilde{q^*}$ | Fleet's minimum forward average SoC (e.g., 30%) |
| | $\omega$ | Vehicle energy efficiency (e.g., 0.248 kWh/mi) |
| | $V_F$ | Number of EV fleet vehicles (e.g., 15,000 SAEVs) |
| | $B$ | Battery capacity per (homogeneous) fleet vehicle (e.g., 90 kWh) |
| | $\widetilde{Q_F}$ | Target average fleet SoC at end of day (e.g., 80%) |
| | $H$ | Time window (hours) for forward average SoC calculation (e.g., 4 hours) |
| | $M$ | Large penalty factor to enforce average fleet end-of-day SoC target (e.g., 2,500,000) |
| | $W(t)$ | Demand-responsive fleet use indicator at time step $t$ |
| | $C(t)$ | Intended charging operational availability indicator at time step $t$ (0 or 1) |
| | $d(t)$ | Aggregate distance (mi) traveled by fleet during time step $t$ |
| | $q_{min}^F$ | Minimum fleet average SoC (e.g., 10%) |
| | $q_{max}^F$ | Maximum fleet average SoC (e.g., 100%) |

The objective function, shown in Eq. (1), minimizes the operational opportunity cost of charging while meeting sufficient battery energy for continued fleet. A time-varying fleet-use factor, $W(t)$, is defined as the average duration that a fleet vehicle is actively serving ride-hailing requests during time step *t*. This measure reflects the interaction between passenger demand and fleet supply and captures the opportunity cost of temporarily removing vehicles from service for charging. A larger $W(t)$ indicates higher fleet use at time step *t*, and therefore a higher opportunity cost of charging. Weighting charging energy by $W(t)$ encourages charging towards lower fleet use hours, while preserving vehicle availability during periods of high passenger demand.

To ensure sufficient energy reserves for continued multi-day fleet operations, the model penalizes deviations from a predefined target fleet-average SoC at the end of the day $\widetilde{Q_F}$. A large penalty factor, $M$, is applied to the absolute difference between the predicted end-of-day fleet-average SoC, $\frac{Q_F(T+1)}{V_F B}$, and $\widetilde{Q_F}$, thereby ensuring sufficient energy reserves for the following operating day.

$$\min \sum_{t=1}^{T} E_b^F(t)W(t) + \left|\widetilde{Q_F} - \frac{Q_F(T+1)}{V_F B}\right| M \tag{1}$$

$S.t.$

$$Q_F(t+1) = Q_F(t) + E_b^F(t) - \omega d(t), \forall t \in T \tag{2}$$

$$q_{min}^{F} \leq \frac{Q_F(t)}{V_F B} \leq q_{max}^{F}, \forall t \in T \quad (3)$$

$$\widetilde{q^*} \leq \frac{\sum_{h=1}^{H} \frac{Q_F(t+h)}{V_F B}}{H}, \forall t \in T \quad (4)$$

$$0 \leq E_b^F(t) \leq E_b^{Fmax}(t) C(t), \forall t \in T \quad (5)$$

$$Q(\cdot),\ E_b^F(\cdot), E_s^F(\cdot), E_s^{HH}(\cdot) \in \mathbb{R}^+ \quad (6)$$

The aggregate fleet battery state is updated at each time step based on mobility-related energy consumption. For a homogeneous fleet, total battery capacity $Q_F(t)$, is given by the product of fleet size, $V_F$, and individual vehicle battery capacity, $B$. As shown in Eq. (2), stored fleet energy at time step *t+1* equals the stored energy at time t, plus scheduled charging, $E_b^F(t)$, minus energy consumed by fleet travel. Travel energy consumption is estimated from aggregate fleet distance traveled, $d(t)$, obtained from POLARIS, and the assumed vehicle energy-consumption rate, $\omega$ (in kWh/mile).

Fleet-average SoC is constrained within predefined minimum and maximum bounds in Eq. (3). A forward-looking minimum SoC constraint, Eq. (4), further requires the fleet-average SoC over a specified future time window $H$, $\frac{\sum_{h=1}^{H} \frac{Q_F(t+h)}{V_F B}}{H}$, to remain above a predefined threshold, $\widetilde{q^*}$. This constraint maintains sufficient energy reserves for expected fleet activity and prevents charging schedules from compromising subsequent fleet service.

Charging feasibility is further constrained by a time-dependent operational availability indicator, $C(t)$. This parameter indicates whether charging is operationally available during time step t, where $C(t) = 0$ represents periods when charging is operationally unavailable (e.g., overnight hours when charging facilities are normally unstaffed for manually connecting SAEVs to chargers), whereas $C(t) = 1$ indicates full charging availability. The maximum charging energy at each time step is therefore jointly determined by the supplied EVCS configuration (including available plug counts and charging power) and $C(t)$.

Together, these components enable charging schedules to adapt to temporal fleet use and practical charging availability without relying on grid signals.

### 2.3 Discharge Potential from Personal EVs

Before incorporating personal/HH EVs into the day-ahead optimization framework, their maximum feasible discharge potential at each hourly time step is quantified. Following the eligibility criteria of Metcalfe et al. (2025), only HH EVs that are either already parked at home or have a scheduled activity destination of home, with access to home chargers and equipped with V2G-compatible hardware, are eligible to participate in V2G operations. In addition, an EV's SoC must exceed a predefined threshold to ensure basic mobility needs are preserved. Variables used to identify eligible HH EVs for V2G participation and to quantify their discharge potential are summarized in Table 2, along with example assumptions.

Table 2. Variables for V2G Eligibility and Discharging Potential of HH EVs

| Symbol | Description |
|---|---|
| $t$ | Simulation time step, $t \in [0, T]$ |
| $I(t)$ | Set of personal EVs eligible for V2G at time step *t*; $i \in I$ |
| $E_{is}^{max}(t)$ | Individual maximum discharge capacity of HH EV *i* during time step *t* |
| $SoC_i(t)$ | SoC of HH EV $i$ at the beginning of time step *t* (e.g., 40%) |
| $SoC^{min}$ | Minimum allowable SoC for V2G discharge (e.g., 30%) |
| $P_i$ | Discharge power of HH EV i's home charger (e.g., 7 kW) |
| $\Delta t$ | Expected duration of the next discharge interval (e.g., 0.25 hour) |
| $B_i$ | Battery capacity of HH EV $i$ (e.g., 74 kWh) |
| $E_{Is}^{max}(t)$ | Aggregate HHEV discharge capacity (kWh) at time step *t* |

The set of eligible EVs at each time step is denoted as $I(t)$. For each eligible personal EV $i \in I(t)$ at time step *t*, the maximum feasible energy (kWh) available for discharging over $\Delta t$ time interval, $E_{is}^{max}(t)$, is calculated as Eq. (7). It takes the minimum of two values: the energy deliverable at the home charger's discharge power rate $P_i$ over $\Delta t$ period, and the usable battery energy, given by the product of EV's battery capacity (in kWh) $B_i$ and the difference between EV's current $SoC_i(t)$ and a predefined minimum SoC allowed for discharging, denoted as $SoC^{min}$. This ensures the discharge is limited by both charger power constraints and available battery energy above the minimum SoC threshold.

$$E_{is}^{max}(t) = \min\{P_i\, \Delta t, \left(SoC_i(t) - SoC^{min}\right) B_i\} \tag{7}$$

The aggregate discharging potential across all participating HH EVs at time step *t*, denoted $E_{Is}^{max}(t)$, is then obtained by summing the individual discharge capacities, as shown in Eq. (8), yielding the aggregate maximum feasible energy to sell from personal EVs.

$$E_{Is}^{max}(t) = \sum_{i \in I} E_{is}^{max}(t) = \sum_{i \in I} \min\{P_i\, \Delta t, \left(SoC_i(t) - SoC^{min}\right) B_i\} \tag{8}$$

This aggregate value represents the total feasible discharging energy from participating HH EVs and serves as an upper bound in the following smart charging optimization model.

### 2.4 Coordinated Day-ahead Smart Charging Model

This study integrates the day-ahead optimization framework developed in Su et al. (2026) with the POLARIS agent-based simulation platform to model coordinated charging for the commercial EV fleet and other EVs, including firm-owned and personal ones. In this work, HH EVs are used as representative personal EVs within POLARIS, although the framework can be extended to various privately owned EV types.

The integrated model operates on a rolling 24-hour planning horizon. At each simulation hour, time-varying inputs (such as electricity prices, pollutant intensity, grid load, and grid committed capacity) are updated so that the current hour becomes the start of the next 24-hour window. The optimization module then re-solves energy transaction schedules in this horizon at each 15-minute timestep, using the most recent simulated fleet and HH EV states. The objective is to minimize the total energy-related costs over the entire simulation day while maintaining fleet service reliability. Cost components include fleet electricity cost, health-damage cost, V2G revenues earned by both fleet and HH EVs, battery degradation costs, and large penalties for deviations from the target end-of-day fleet SoC to support multi-day operational continuity. For example, when the simulation reaches 6 AM, the model optimizes hourly energy decisions for the subsequent 24-hour period, through 6 AM the following day, aligning (dis)charging decisions with anticipated operational needs.

These models enable flexible evaluation of several charging strategies, including DAC, V1G, fleet-only V2G (Fleet-V2G), and Combined-V2G involving both fleet and HH EVs. This framework creates a two-way feedback loop in which optimization informs EV behavior and activity plans, while the simulation captures operational dynamics such as travel demand, network congestion, and spatial and temporal vehicle availability, and feeds the resulting system state back to the optimization step. Compared to methods that assume perfect adherence to optimized schedules and do not consider traffic or activity patterns, the proposed model captures dynamic feedback between optimization decisions and simulated system conditions, reflecting more realistic adjustments driven by dynamically simulated fleet conditions. Figure 1 shows the overall workflow.

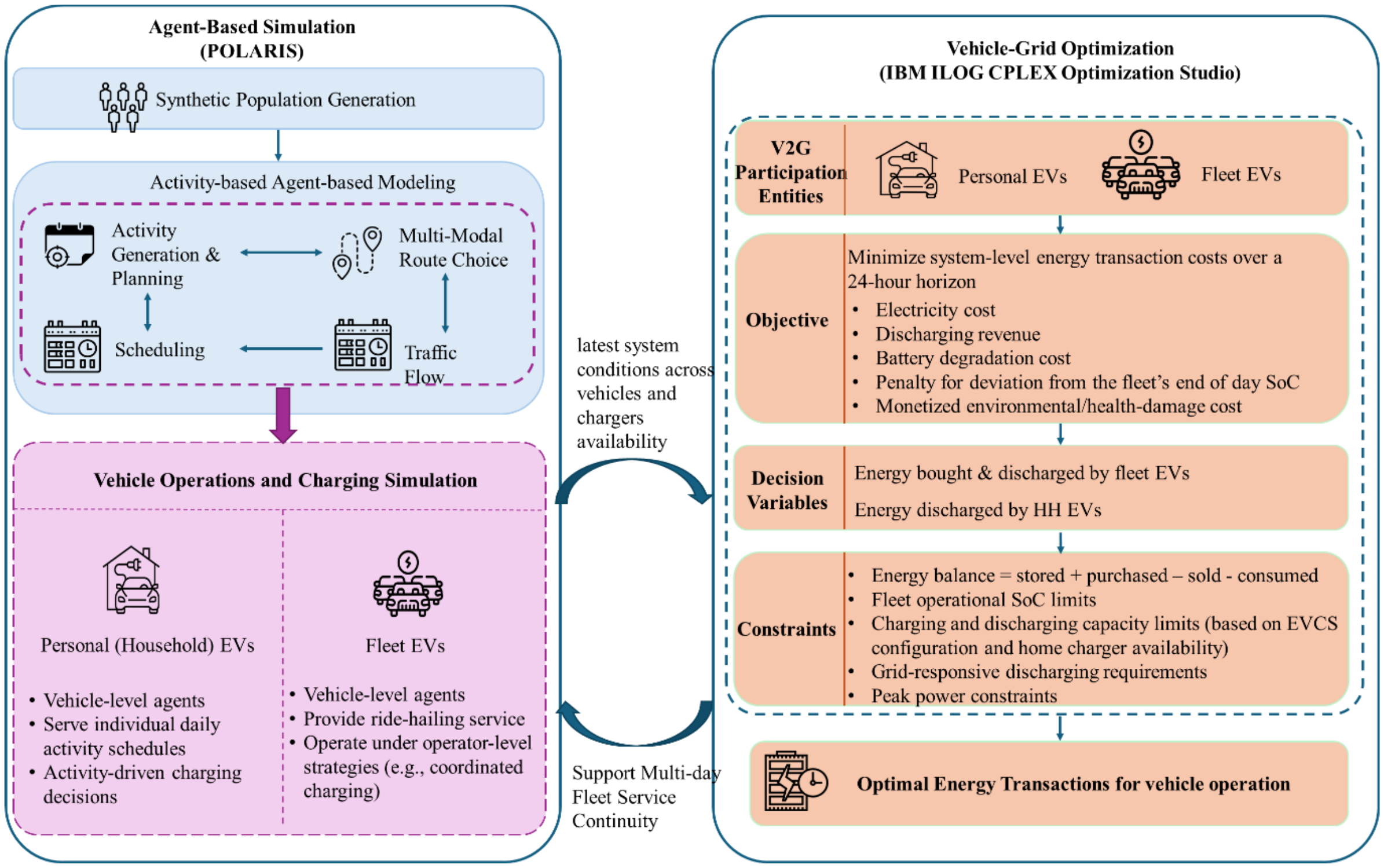

Figure 1. Simulation-Optimization Framework for the Coordinated Charging Model

The smart-charging optimization determines three energy-transaction decision variables: fleet charging and discharging energy, and aggregated HH EV discharging energy. Notably, HH EV charging is not modeled as a decision variable, reflecting the reality that utility operators typically lack the authority to control residential charging behavior driven by individual commuting needs and home charging preferences. To implement the optimized energy transactions, a within-day dispatch optimization model from Dean et al. (2023) is adopted to plan and assign tasks to fleet vehicles, including (dis)charging, maintenance, passenger service, and repositioning.

## 3. CASE STUDY

Austin, Texas, is selected as the case study area because of its rapid EV adoption, active development of automated EV fleets, and supportive charging infrastructure network. Austin is leading the statewide transition to electric mobility, accounting for roughly 19.13% of all EVs registered in Texas and holding the highest EV share among major Texas metropolitan areas, with 5.13% of all registered vehicles being electric as of December 2025 (Dallas-Fort Worth Clean Cities, 2025). The region continues to advance its electrification goals through the Austin Climate Equity Plan, which targets 40% of all VMT to be electric by 2030 (City of Austin, 2020). Support for this transition is reflected in the city's expanding public and private charging-network development and local utility incentive programs. According to ChargePoint's national scoring of EV-supportive regions, which considers population, EV adoption, and available EVCS, ranks Austin 4th nationwide, following the San Francisco Bay Area, Seattle, and San Diego (ChargePoint, 2025). Austin Energy, the region's primary utility, offers rebates for charger purchase or installation, and runs the Power Partner EV program, which enables the utility to remotely adjust enrolled Level 2 chargers or qualified vehicles during periods of high system demand to help manage grid loads (Austin Energy, 2025).

Austin has also emerged as an early deployment city for automated electric ride-hailing services. By the end of 2025, Waymo operated over 100 SAEVs across a roughly 90-square-mile service area, while Tesla launched a pilot robotaxi fleet of about 30 vehicles with plans to expand to other metropolitan markets (Campbell, 2025b; Lambert, 2025). These conditions support Austin's suitability for examining integrated EV charging and grid-interactive operations.

Within this context, this study models regional travel activity across the six-county Austin metropolitan region, covering approximately 5,300 square miles with 2,160 zones. This large-scale configuration provides a representative setting for evaluating coordinated (dis)charging and grid-interactive behaviors of personal EVs and EV fleets under time-varying electricity market and grid conditions.

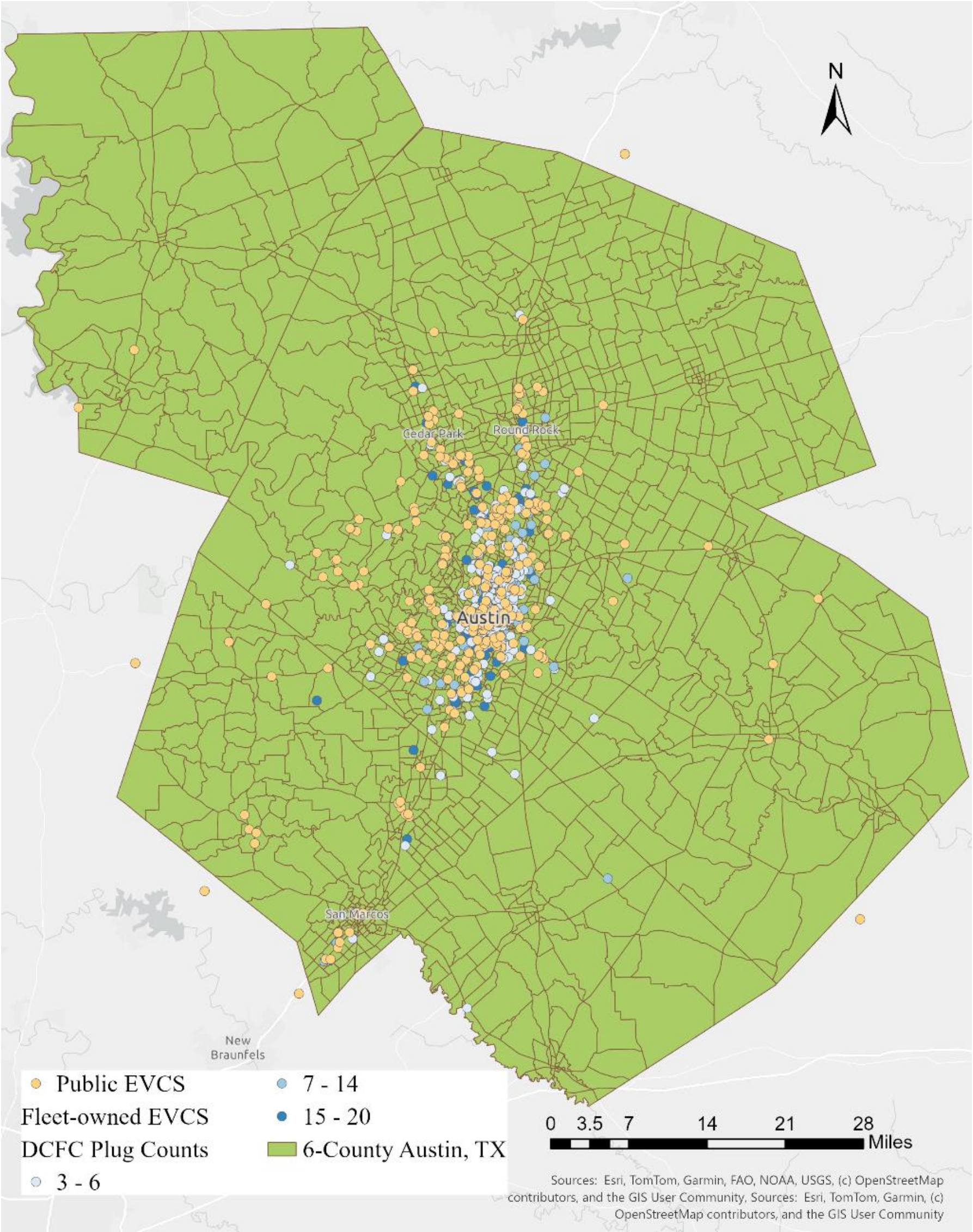

Figure 2. Austin 6-county Region, TX, along with EVCS distribution

A ride-hailing fleet of 15,000 homogeneous SAEVs is simulated within the Austin 6-county region to serve traveler demand and ride-hailing requests. Each SAEV is equipped with a 90-kWh battery that provides about a 300-mile range at full capacity. To account for operational reserve needs and battery degradation, the operational SoC is restricted between 10% and 100%. This yields an effective usable range of around 270 miles at full charge. The ride-hailing fare structure includes a $2 base pickup fee, $0.90/mile, and $0.25/minute. To encourage ride-sharing, travelers who opt for shared rides receive a 40% fare discount on the $0.54 per-mile and $0.15 per-minute charges.

The SAEV fleet and personal EVs are supported by a strategically deployed charging-station network that integrates both public and fleet-owned/private charging infrastructure. Prior studies have highlighted the benefits of a public-private (PP) EVCS strategy (Su et al., 2024; Su et al., 2025), which opens fleet-owned charger access for broader EV users. This approach enables charger-sharing across user groups and makes the fleet eligible for public incentives, thus improving charger use, lowering required investment, and reducing charging delays. Figure 2 shows the spatial distribution of all public EVCS based on Alternative Fueling Station Locators (Department of Energy, 2025c), along with the sited PP EVCS serving personal EVs and the SAEV fleet within the study region. On average, the Austin 6-county region contains approximately one PP EVCS per 10 square miles, with each station housing about 7.8 DCFC plugs rated at 120 kW. The SAEV fleet is assumed to rely exclusively on DCFC chargers for both charging and discharging to support high-use and reduce charging downtime. This assumption aligns with the industry's emphasis on minimizing downtime to remain competitive in the ride-hailing market (NovaCharge, 2024). Personal EVs, in contrast, can use either DCFC or Level 2 chargers at the charging stations or Level 2 home chargers; however, their discharging activities are limited to home chargers. Only HH EVs with access to home charging are considered eligible for participation in fleet-plus-household V2G operations.

Given that V2G adoption typically requires costly charging infrastructure upgrades and advanced bidirectional chargers, which can limit large-scale household participation, this study assumes a 40% participation rate based on Li et al. (2024). Together, these HH EVs and EV fleets form the set of grid-interactive entities evaluated in the smart charging and related operational behaviors within the study area.

### 3.1 Electricity Market and Grid Conditions

The Electric Reliability Council of Texas (ERCOT) manages the only fully independent electricity grid in the United States and coordinates energy supply for most of Texas. It covers roughly 90% of the state's electric demand and includes major metropolitan areas such as Austin, San Antonio, Houston, and Dallas-Fort Worth (ERCOT, 2025a). For wholesale market settlement, ERCOT is divided into multiple load zones, including Austin Energy's service area.

Within this system, the analysis incorporates time-varying electricity prices to capture the economic conditions relevant to EV charging and V2G operations. The time-of-use (TOU) fare structure is referred to the Austin Energy Electric Tariff (City of Austin, 2025), which uses a lower rate of \$26.77/MWh during late-night and early-morning hours (10 PM to 7 AM), and a mid-peak rate of \$41.18/MWh during morning and afternoon (7 AM to 5 PM), and an evening peak rate of \$140/MWh between 5 PM and 10 PM.

To characterize real-time pricing, hourly locational marginal prices (LMPs) for the Austin Energy load zone were obtained from ERCOT from November 2024 through October 2025 to enable a year-round analysis (GridStatus, 2025a). Following classification methods used in prior studies (Zhang and Chen, 2020; Dean et al., 2023; Su et al., 2025), hourly LMP observations were sorted into four categories based on price variability and deviation patterns: Peak, No-Peak, Off-Peak, and Spike, with corresponding shares of 61.3%, 6.2%, 6.2%, and 26.3%, respectively. This grouping highlights the diversity of real-time electricity prices in ERCOT and enables the model to reflect both normal and extreme system conditions. Figure 3 shows each LMP category, with navy blue curves indicating daily LMP by hour under that category and a red curve representing the category-level average.

System-level grid conditions were evaluated using ERCOT's real-time grid load and committed generation capacity records for the same period, ensuring consistency with the LMP datasets (GridStatus, 2025b & 2025c). Since ERCOT's system-level data are significantly larger in scale than the city-level setting of this study, grid capacity and load were rescaled to reflect the grid conditions at the City of Austin. This scaling is based on the ratio between Austin Energy's peak demand record of 3.07 GW in August 2023 and ERCOT's statewide peak of 85.508 GW during the same period (Arnim et al., 2024; ERCOT, 2025b). Planning reserve margins are commonly used to ensure planned generation and capacity are sufficient to meet electricity demand, and recommendations from the North American Electric

Reliability Corporation (NERC) generally set this buffer between 10% and 20% (Lawson, 2022). However, ERCOT's own reserve forecasts indicate its reserve needs are often above the typical planning range, such as a forecasted margin of 22.2% for summer 2023, 29.4% for summer 2024, and 46.7% for summer 2025 due to delayed resource support (ERCOT, 2022; ERCOT, 2023). To reflect these planning considerations and to account for the increasing electricity demand introduced by scalable EV fleet operations, this study uses a 21% reserve-margin threshold to identify grid-stress conditions. This ensures that V2G activation aligns with the grid-stress periods, supporting the grid while respecting vehicles' availability and flexibility to meet personal planned daily travel schedules.

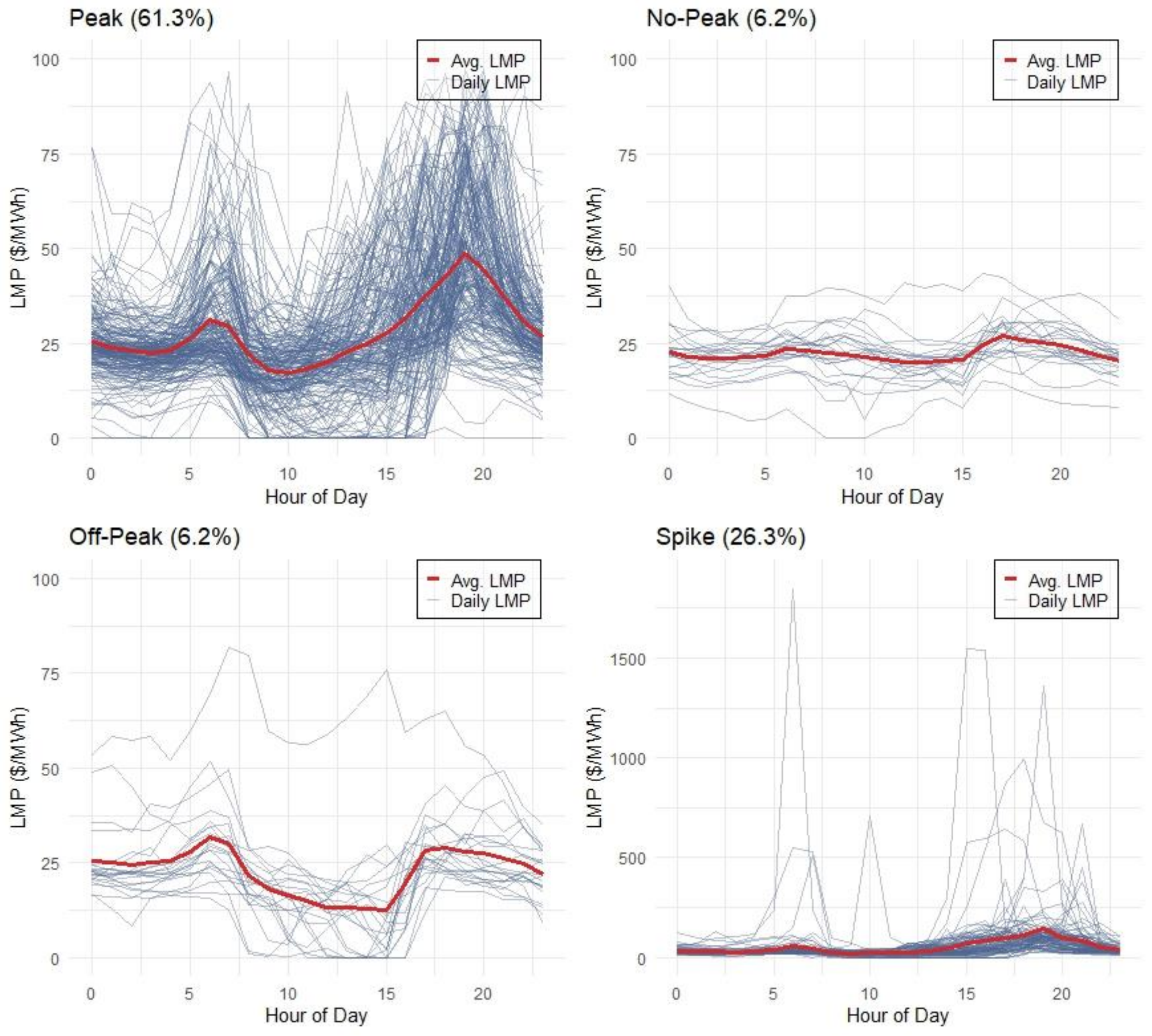

Figure 3. Categorized Hourly LMP Tariffs for the ERCOT Austin Energy Load Zone

### 3.2 Pollutant Data

Power-sector pollutant outputs are evaluated using publicly available, high-resolution datasets. Hourly average emission factors (in lb/MWh) for $CO_2$, $NO_x$, and $SO_2$ in the ERCOT region are obtained from a dataset developed by the National Renewable Energy Laboratory (2023) (recently renamed the National Laboratory of the Rockies) and compiled from federal sources, including the Environmental Protection Agency. These data provide regionally resolved, hourly profiles of generation-related pollutant intensities.

To monetize pollutant-related externalities within the optimization framework, this study adopts representative marginal damage valuations reported by Dean et al. (2023). Pollutant-specific damage costs per metric ton are set at $140/ton for $CO_2$, $4,750/ton for $NO_x$, and $16,600/ton for $SO_2$, based on recent estimates from the U.S. Department of Transportation (2025) and Rhodes (2023). These values are used to translate pollutant intensity factors into corresponding economic cost signals, enabling the evaluation of system-level economic outcomes and associated environmental and health-related impacts.

# 4. RESULTS AND DISCUSSIONS

## 4.1 SAEV Fleet Performance under Alternative Charging Strategies

The impacts of alternative charging strategies on SAEV fleet service outcomes are summarized in Table 3. Overall, passenger-service performance improves modestly from price-agnostic DAC and remains broadly comparable under the grid-responsive charging strategies. DAC coordinates charging according to time-varying fleet use and practical charging availability at EVCS without explicitly responding to electricity prices or grid conditions. Under DAC, each SAEV serves an average of 31 passenger-trips per day at a median passenger wait time of nearly 5 minutes, while daily VMT increases to 270 miles and the empty VMT share to 31%.

Grid-responsive charging maintains similarly stable passenger-service performance across the five electricity-price structures. Each SAEV serves roughly 31 passenger trips per day, with median passenger wait times up to 5 minutes. Average daily VMT ranges from 268 to 279 miles per SAEV, with empty VMT shares of about 30% to 33%. For served passenger trips, median travel distances remain around 9 miles, and average fares remain about $11 per ride. Consequently, daily passenger-service revenues remain within a narrow range of $329 to $333/SAEV, compared with $326/SAEV under DAC. These findings reflect the simulation-optimization framework design, where charging and discharging decisions remain constrained by operational requirements, ensuring that energy transactions do not disrupt mobility services.

Table 3. Fleet Service Performance under Alternative Pricing and Charging Strategies

| Electricity Price | Strategy | Median Passenger Wait Time (minutes) | Avg. Daily Person-trips Served per SAEV | Avg. Daily VMT per SAEV (miles) | Avg. Fleet Empty VMT share (%eVMT) | Median Travel Distance per SAEV Trip (miles) | Avg. Fare per Ride Request ($) | Avg. Daily Passenger-service Revenue per SAEV ($) |
|---|---|---|---|---|---|---|---|---|
| Price-agnostic | Demand-aware Charging (DAC) | 4.97 | 30 | 270 | 31.26 | 8.90 | 10.74 | 326 |
| Spike | V1G | 4.88 | 31 | 279 | 33.21 | 8.93 | 10.75 | 331 |
| | Fleet-V2G | 4.93 | 31 | 279 | 32.81 | 8.99 | 10.76 | 332 |
| | Combined-V2G | 4.65 | 31 | 276 | 31.99 | 8.89 | 10.72 | 333 |
| TOU | V1G | 5.22 | 30 | 271 | 31.27 | 9.07 | 10.83 | 329 |
| | Fleet-V2G | 4.85 | 31 | 269 | 30.31 | 8.97 | 10.76 | 333 |
| | Combined-V2G | 4.82 | 31 | 268 | 30.22 | 8.94 | 10.75 | 332 |
| Peak | V1G | 4.92 | 31 | 279 | 32.67 | 9.00 | 10.79 | 333 |
| | Fleet-V2G | 4.93 | 31 | 279 | 32.81 | 8.99 | 10.76 | 332 |
| | Combined -V2G | 4.65 | 31 | 276 | 31.99 | 8.89 | 10.72 | 333 |
| No-Peak | V1G | 4.90 | 31 | 277 | 32.57 | 9.01 | 10.77 | 332 |
| | Fleet-V2G | 4.97 | 31 | 278 | 32.83 | 9.01 | 10.79 | 332 |
| | Combined -V2G | 5.00 | 31 | 277 | 32.52 | 9.04 | 10.80 | 332 |
| Off-Peak | V1G | 5.07 | 31 | 276 | 32.03 | 9.04 | 10.84 | 333 |
| | Fleet-V2G | 4.87 | 31 | 276 | 32.48 | 8.95 | 10.74 | 330 |
| | Combined -V2G | 4.60 | 31 | 274 | 31.98 | 8.85 | 10.67 | 330 |

**4.2 EV Charging and Discharging Profiles and Grid-support Response**

Hourly EV energy transactions reveal distinct temporal patterns under demand-aware coordination and grid-responsive V1G and V2G operation. Figure 4 presents the temporal distribution of fleet charging (grey bars) and discharging (yellow bars), and HH EV discharging (red bars), along with grid profiles.

DAC explicitly incorporates time-varying fleet use and practical charging availability. No charging occurs before 6 AM, reflecting the assumed overnight EVCS staffing constraint for manually connecting SAEVs to chargers, and only 3% of daily charging occurs between 6 and 9 AM. It then rises sharply after the morning travel peak, with 35% occurring between 10 AM and 2 PM. After declining in the afternoon, charging rebounds between 10 PM and midnight, accounting for another 19%.

Introducing coordinated charging through V1G strategically reshapes the temporal charging pattern by considering electricity prices and grid conditions. Across the five pricing structures, 29% to 42% of daily V1G charging occurs during midday hours (10 AM to 2 PM), while another 18% to 25% concentrates during late-night periods (10 PM to midnight). In contrast, charging from 6 to 9 AM becomes particularly limited (less than 0.2%), compared with 3% under DAC. Unlike DAC, V1G can use early-morning hours when economically attractive, with about 8% of daily charging occurring from midnight to 2 AM under TOU, Peak, and Spike pricing. V1G therefore provides greater flexibility to schedule energy purchases during hours with lower electricity prices and favorable grid conditions.

Transitioning from V1G to Fleet-V2G modifies these temporal patterns by jointly coordinating charging and selective discharging. Midday charging accounts for 32% to 41% of daily fleet energy consumption across the five price structures, reaching approximately 41% under Spike. Fleet discharge is highly selective, occurring primarily under TOU and Spike pricing and concentrating during evening peak periods (7 to 10 PM), when both grid-support needs and economic conditions justify participation. In contrast, fleet discharging under the other pricing structures remains relatively limited, reflecting weaker economic incentives after accounting for battery degradation and operational constraints. These results highlight the role of electricity pricing in shaping both the timing and intensity of V2G participation. Greater intra-day price variation encourages charging during favorable periods while creating stronger incentives for selective discharging during high-price evening hours.

Allowing HH EV participation through Combined-V2G further expands grid support. Relative to Fleet-V2G, the fleet-side charging pattern remains broadly similar, with 29% to 41% of daily fleet charging occurring between 10 AM and 2 PM. As under V1G and Fleet-V2G, charging between 6 and 9 AM remains negligible. Fleet discharging continues to occur primarily under TOU and Spike, while HH EVs provide additional discharge (mainly from 5 to 11 PM) under all five price structures. HH EV participation therefore broadens the temporal availability of V2G resources beyond the more selective fleet response, providing complementary grid support when SAEVs may be constrained by passenger-service obligation. Although HH EV discharge remains relatively small at the individual-vehicle level, aggregating these distributed resources can provide meaningful system-level grid support, particularly as EV adoption and V2G-capable infrastructure expand.

Across the grid-responsive strategies, total daily fleet charging also remains broadly comparable with DAC, indicating that smart charging primarily redistributes energy transactions over time rather than substantially changing overall fleet charging demand. In general, Figure 4 reveals a clear progression in charging and discharging behaviors across the evaluated levels of coordination. DAC establishes an operationally feasible charging schedule around fleet use, with pronounced post-morning-peak and late-night rebound periods. V1G introduces load shifting by reallocating charging away from grid-stress and high-cost periods. Fleet-V2G preserves these coordinated charging patterns while adding targeted fleet-based discharge support during grid-stress periods. Combined-V2G further enhances grid responsiveness by incorporating distributed HH EVs into the response pool, expanding both the capacity and temporal flexibility of discharge support, particularly under electricity prices structure with strong intraday variation.

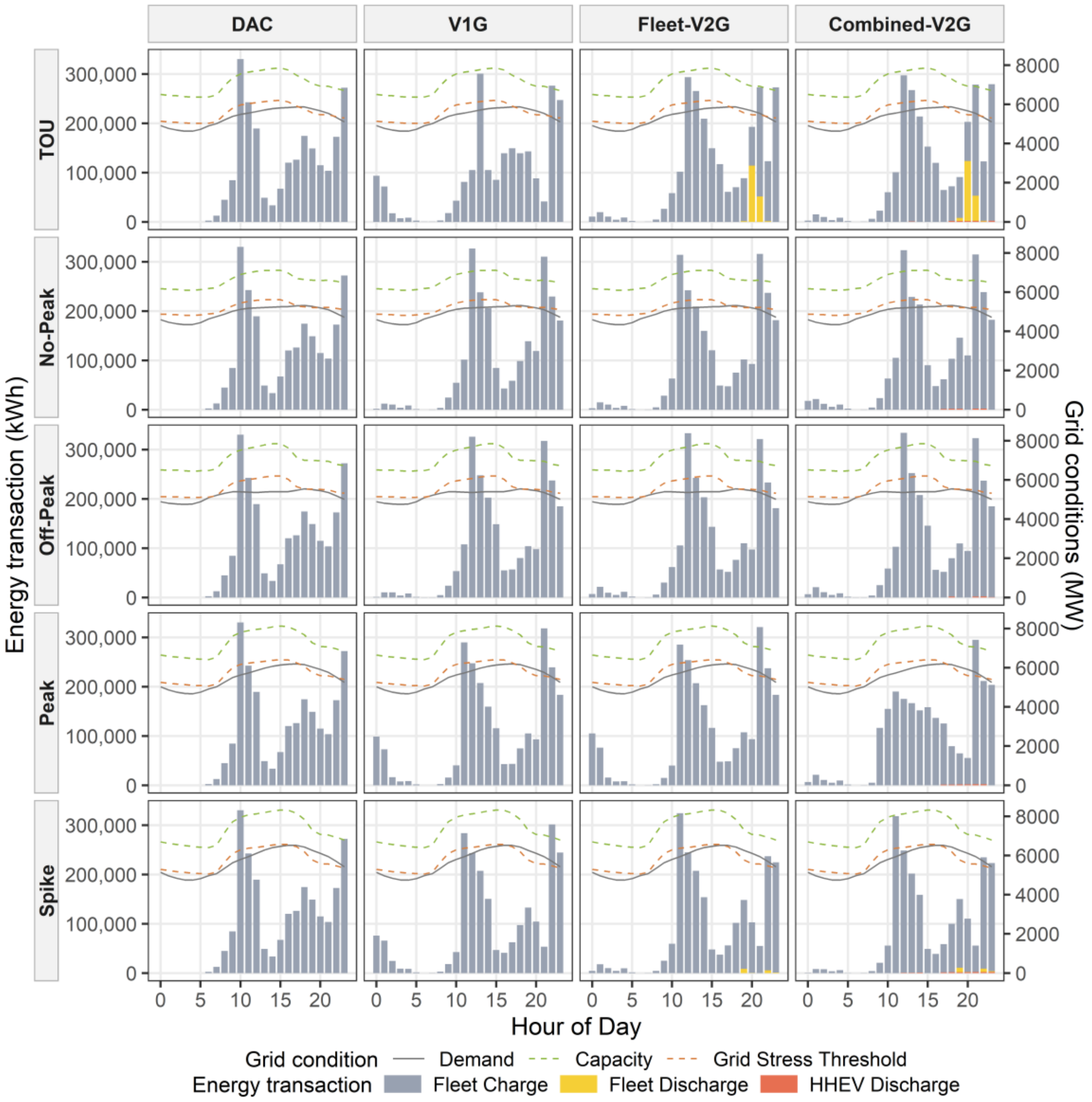

Figure 4. Hourly EV Energy Transactions and Grid Conditions by Charging Strategies and Pricing

### 4.3 Pollutant Outputs and Health-damage Implications

Pollutant outputs and monetized damage costs under alternative charging strategies and five electricity pricing structures are summarized in Table 4. Since differences in fleet activity (e.g., daily VMT per SAEV) are observed across scenarios, environmental impacts are evaluated using both per-SAEV and VMT-normalized emission metrics. HH EV discharge is evaluated at the aggregate level because its per-vehicle effects are small but can become meaningful when aggregated across participating vehicles. In addition to $CO_2$ emissions, monetized $CO_2$ and pollutant-related damage costs (including those associated with $NO_x$ and $SO_2$) are also evaluated to capture broader societal impacts of smart charging strategies.

Under DAC, net emissions reach 45 kg $CO_2$/SAEV/day, with an intensity of 0.17 kg/VMT and total monetized damages of roughly $10/SAEV/day. Grid-responsive V1G produces modest and highly tariff-dependent changes relative to DAC. Under Spike and TOU pricing, net daily $CO_2$ emissions fall below 44 kg/SAEV/day, with intensities declining to 0.16 kg/VMT. In contrast, V1G under the flatter No-Peak and Off-Peak pricing signals yield little change in daily emissions, despite small reductions in emission intensity, while Peak slightly increases per-SAEV emissions. These results indicate that the environmental benefits of smart charging depend on whether economically favorable charging periods align with lower-emission and lower-stress grid conditions. With pronounced intraday price variation (such as Spike and TOU), V1G concentrates charging in midday periods that often coincide with lower grid stress and greater renewable energy availability, thereby reducing emissions.

Fleet-only V2G provides additional environmental benefits through strategically timed bidirectional discharge when pricing structures provide sufficient incentives for discharge. Under Spike and TOU, net daily $CO_2$ emissions fall below 40 kg/SAEV/day, with reductions of up to approximately 14% relative to DAC in both per-SAEV net emissions and emission intensity. Comparing gross and net $CO_2$ emissions further reveals distinct pathways to emissions reduction. Spike achieves most of its benefit through coordinated charging, with limited additional savings from V2G discharge, whereas TOU relies more strongly on bidirectional discharge to offset higher gross charging emissions. Thus, similar net outcomes can arise from different charging and discharging patterns. Fleet-V2G provides little additional reduction under the remaining flatter tariffs. These results show that the environmental value of V2G is strongly tariff-dependent and increases when discharge incentives coincide with periods of higher grid stress and emissions.

Allowing HH EV participation through Combined-V2G expands the discharge capacity and temporal flexibility while preserving the observed fleet-level environmental benefits under Fleet-V2G. Relative to Fleet-V2G, Combined-V2G generally produces similar or slightly lower per-SAEV and VMT-normalized emissions, which generally remain below V1G levels. This suggests that much of the fleet-side environmental benefit is already captured through coordinated fleet V2G operation.

HH EV participation provides additional system-level pollutant savings, particularly under electricity prices with strong intra-day variation. The largest aggregate $CO_2$ savings occur under Spike pricing, for which HH EV contributes saving roughly 5,196 kg/day of $CO_2$, followed by TOU (3,701 kg/day) and Peak pricing (3,683 kg/day). These results suggest that distributed HH EV participation becomes more effective when price signals provide strong incentives for discharging. Under flatter and lower-value price signals, weaker arbitrage opportunities are insufficient to offset V2G-related costs, limiting HH EV participation and the associated air-quality benefits.

The monetized damage results show a similar tariff-dependent pattern. Under Spike and TOU, grid-responsive strategies reduce monetized damages relative to DAC, while Fleet-V2G and Combined-V2G yield comparable fleet-level damages but generally provide further reductions relative to V1G.

The above results highlight that environmental gains depend jointly on charging flexibility and the temporal alignment of electricity-price signals with grid and emissions conditions. Relative to DAC, V1G provides modest improvements when price signals favor lower-impact charging periods, while Fleet-V2G further reduces emissions under Spike and TOU through strategically timed discharge. Combined-V2G further broadens this response through distributed HH EV participation, translating small vehicle-level contributions into meaningful aggregate $CO_2$ savings.

Table 4. Carbon Impacts Across Charging Strategy and Electricity Price

| Electricity Pricing | Strategy | Daily $CO_2$ per SAEV (kg) | Daily $CO_2$ Reduction per SAEV (kg) | Net Daily $CO_2$ per SAEV (kg) | Net $CO_2$ Intensity per SAEV (kg/VMT) | Daily $CO_2$ Savings by HHEV (kg) | Daily Net $CO_2$ Damage Cost per SAEV ($) | Daily Net Pollutant Damage Cost per SAEV ($) |
|---|---|---|---|---|---|---|---|---|
| Price-agnostic | DAC | 45.18 | 0.00 | 45.18 | 0.17 | 0.00 | 6.32 | 3.39 |
| Spike | V1G | 43.78 | 0.00 | 43.78 | 0.16 | 0.00 | 6.13 | 3.35 |
| | Fleet-V2G | 40.28 | 0.35 | 39.93 | 0.15 | 0.00 | 5.59 | 3.03 |
| | Combined-V2G | 39.50 | 0.39 | 39.11 | 0.15 | 5195.53 | 5.47 | 2.96 |
| TOU | V1G | 43.55 | 0.00 | 43.55 | 0.16 | 0.00 | 6.10 | 3.26 |
| | Fleet-V2G | 42.90 | 3.95 | 38.96 | 0.14 | 0.00 | 5.45 | 3.01 |
| | Combined-V2G | 42.99 | 4.24 | 38.76 | 0.14 | 3700.61 | 5.42 | 2.99 |
| Peak | V1G | 46.39 | 0.00 | 46.39 | 0.17 | 0.00 | 6.49 | 3.49 |
| | Fleet-V2G | 45.57 | 0.00 | 45.57 | 0.16 | 0.00 | 6.38 | 3.41 |
| | Combined-V2G | 45.15 | 0.00 | 45.15 | 0.16 | 3683.45 | 6.32 | 3.41 |
| No-Peak | V1G | 45.08 | 0.00 | 45.08 | 0.16 | 0.00 | 6.31 | 3.32 |
| | Fleet-V2G | 44.99 | 0.00 | 44.99 | 0.16 | 0.00 | 6.30 | 3.37 |
| | Combined-V2G | 46.24 | 0.00 | 46.24 | 0.17 | 3218.36 | 6.47 | 3.42 |
| Off-Peak | V1G | 45.23 | 0.00 | 45.23 | 0.16 | 0.00 | 6.33 | 3.36 |
| | Fleet-V2G | 45.75 | 0.00 | 45.75 | 0.16 | 0.00 | 6.40 | 3.41 |
| | Combined-V2G | 45.78 | 0.00 | 45.78 | 0.16 | 2091.57 | 6.41 | 3.41 |

### 4.4 Fleet-level Economic Implications

Daily per-SAEV economic outcomes under alternative charging strategies are summarized in Figure 5, including daily electricity cost, passenger-service operating cost and profit, and overall fleet-operator profit. Passenger-service profit is calculated as passenger-service revenue minus operating cost, while fleet-operator profit additionally accounts for electricity costs, V2G discharging revenues, and battery degradation costs associated with V2G participation.

DAC yields an average fleet-operator profit of approximately $209/SAEV/day, with grid-responsive strategies slightly increasing both average passenger-service profit and fleet-operator profit. V1G consistently increases fleet-operator profit by about 2% ($4/SAEV/day) across five pricing structures, while electricity costs remain broadly comparable to DAC. The improvement is driven primarily by higher passenger-service profit, suggesting that the economic value of grid-responsive charging stems mainly from improved coordination between charging and passenger-service operations rather than direct electricity-cost savings.

The incremental economic value of Fleet-V2G is smaller. Relative to V1G, Fleet-V2G produces its largest gain under TOU pricing ($5/SAEV/day, or 2.4%), while changes under the other pricing structures are limited. Expanding participation through Combined-V2G provides similarly modest changes relative to Fleet-V2G.

Overall, the economic results indicate that fleet profitability is primarily shaped by passenger-service profit. Grid-responsive charging provides modest economic improvements over DAC, while the incremental value of V2G depends on the magnitude and temporal alignment of electricity-price signals with fleet operations.

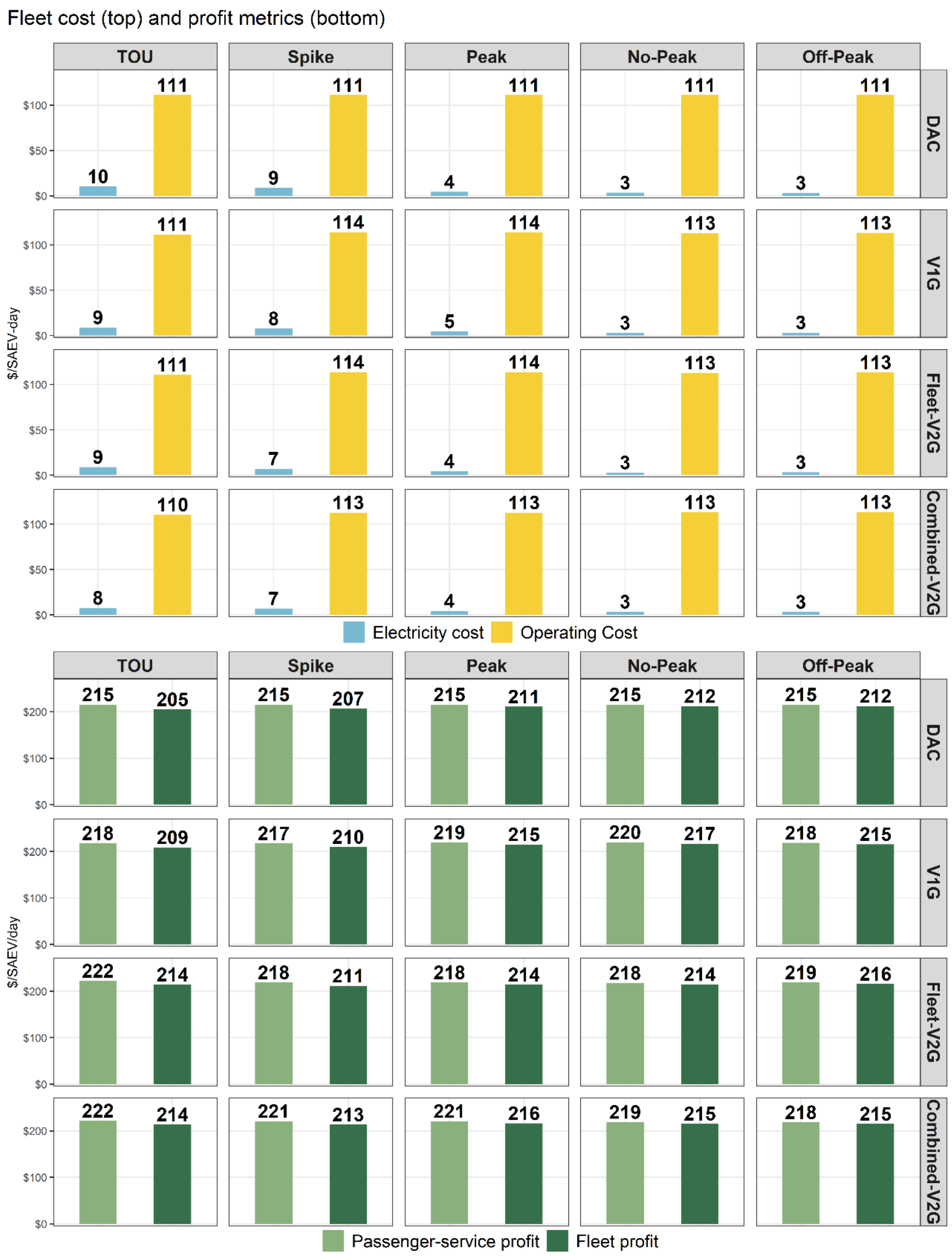

Figure 5. Per-SAEV Daily Economic Components Across Electricity Prices and Charging Strategies

Note:

[1] Daily passenger-service profit per SAEV is calculated as: Daily passenger-service revenue per SAEV - Daily operating cost per SAEV

[2] Daily Profit per SAEV = Daily passenger-service profit per SAEV- Daily electricity cost per SAEV + Daily discharging revenue per SAEV – Daily battery discharging degradation cost per SAEV

### 4.5 Comparative Implications of Coordinated Charging Strategies

The four charging strategies represent different levels of operational charging coordination and grid interaction, each offering distinct advantages depending on technical readiness, tariff structures, and EV participation scale. Across the evaluated scenarios, DAC demonstrates that benefits can be achieved through improved fleet-side charging coordination, while grid-responsive V1G and V2G strategies extend this coordination to account for electricity prices, grid-stress conditions, and pollutant-related damage costs. All coordinated strategies maintain broadly comparable ride-hailing service outcomes.

Conventional uncontrolled charging follows immediate vehicle energy needs with limited coordination of charging timing. DAC provides a more operationally efficient alternative by scheduling charging around time-varying fleet use and practical charging availability. This allows charging to be shifted toward periods when more vehicles can be charged without substantially affecting passenger service. Importantly, DAC achieves these benefits primarily through improved scheduling and fleet coordination, without requiring bidirectional charging capability or major upgrades to existing charging infrastructure. The results therefore position DAC as a practical and relatively low-cost operational strategy for SAEV fleets, particularly where advanced grid-responsive charging systems or V2G infrastructure are not yet available.

Compared with DAC, V1G more strongly redistributes charging toward lower-cost and lower-stress periods and provides direct grid-responsive flexibility while maintaining similar fleet-service performance. Since V1G does not require bidirectional charging infrastructure, it also represents a practical near-term strategy for large-scale smart charging deployment when V2G infrastructure is limited or costly.

In contrast, Fleet-V2G and Combined-V2G require additional bidirectional charging capability, resulting in higher upfront costs than conventional chargers. Although the incremental vehicle-side upgrade for V2G compatibility is relatively modest, typically estimated at $200 to $400 per vehicle for onboard power electronics and control modules (Steward, 2017), the primary economic barrier is associated with V2G-capable charging infrastructure. As of 2025, a bidirectional Level 2 charger typically costs about $1,200 to $2,500, while installation, permit, utility interconnection equipment, and electrical upgrades can increase total installed costs to roughly $1,700 to $4,000 per charger (Shop Energy Prices, 2025). Although these costs are expected to decline as EV adoption expands and charging technologies mature, early deployment may remain constrained in the near term.

For fleet operators, V2G deployment often involves higher-power DCFC infrastructure capable of supporting large-scale fleet (dis)charging operations. Conventional 50-kW DCFC systems commonly cost around $4,000 to $10,000 per charger, while V2G-capable DCFC chargers can increase costs to between $5,000 and $15,000 per charger depending on charging power, interconnection requirements, and communication modules (Shop Energy Prices, 2025). For comparison, standard residential Level 2 chargers typically cost about $400 to $800 per unit, whereas bidirectional V2G chargers generally cost $1,200 to $2,500, with total costs ranging from approximately $1,700 to $4,000 after accounting for permit, installation, and utility interconnection equipment (Shop Energy Prices, 2025). Fleet operators deploying large numbers of V2G chargers may benefit from economies of scale through bulk-purchase discounts (approximately 10% to 20%) and streamlined installation processes, reducing total installation costs to roughly $1,100 to $2,000 per charger for large fleet installations (Shop Energy Prices, 2025). Fleet-V2G becomes more practical once bidirectional equipment is available within fleet charging facilities, where centralized control enables coordination of vehicle availability with grid-response needs. This situation may resemble the early-stage charging infrastructure deployment, where higher upfront costs are initially absorbed by charging network providers or fleet operators before widespread household adoption. Fleet discharge during strategically selected hours provides grid-responsive flexibility and modest revenues beyond the economic gains achieved through V1G.

These cost considerations help explain the distinct practical roles of the evaluated charging strategies. V1G represents the most immediately deployable and scalable smart charging strategy because it requires minimal infrastructure upgrades while already capturing most of the realized economic and environmental benefits observed in this study. Fleet-V2G becomes increasingly attractive as bidirectional DCFC infrastructure matures within centralized fleet charging facilities, where coordinated vehicle scheduling and higher charger use can better justify additional investment costs. Combined-V2G offers the broadest long-term flexibility and distributed grid-support potential by incorporating HH EV participation, particularly under tariffs with strong intraday price variation. As EV ownership expands and residential chargers become more advanced and widespread, broader participation from both fleet and HH EVs through Combined-V2G may unlock larger system-level benefits by combining centralized fleet coordination with distributed residential energy support. However, its large-scale deployment likely depends on continued reductions in bidirectional charger costs, wider availability of V2G-compatible vehicles, and stronger utility incentive programs that improve participation economics for HH EV owners. Overall, the results suggest a gradual pathway toward increasingly coordinated charging.

## 5. CONCLUSION

This work examines how coordinated charging strategies can support electric mobility systems and power-grid operations as transportation becomes increasingly electrified and automated. To address the limited integration of dynamic mobility operations in existing studies, this research develops an integrated simulation-optimization framework that embeds distinct demand-aware charging and rolling day-ahead smart-charging models within an agent-based simulation environment. Using the Austin 6-county region as a case study, the analysis evaluates DAC, V1G, Fleet-V2G, and Combined-V2G in terms of fleet operations, temporal energy demand, grid-support behavior, environmental impacts, and fleet-level economic outcomes across alternative electricity-price structures.

DAC identifies distinct charging windows around anticipated fleet use and practical charging availability at EVCS, thereby limiting SAEV withdrawals during higher-demand periods. Grid-responsive V1G further reallocates charging toward lower-cost and lower-stress periods, particularly midday and late-night hours, while maintaining passenger wait times, trips served, and passenger-service revenues comparable to those under DAC. These findings position DAC as a practical intermediate strategy between reactive SoC-based charging and grid-responsive smart charging.

V2G offers additional grid and environmental value, but these gains are selective and strongly dependent on the alignment among electricity prices, grid-stress periods, pollutant intensities, and vehicle availability. HH EV participation broadens the temporal and aggregate discharge potential, particularly when SAEVs are constrained by passenger-service needs.

Several modeling assumptions provide opportunities to extend the proposed framework. The current DAC formulation uses historical fleet-use patterns to capture time-varying vehicle availability. Although this approach provides a more operationally realistic baseline, historical patterns may not adequately represent unexpected demand surges, special events, severe weather, or day-to-day changes in travel behavior. Future research could incorporate probabilistic day-ahead forecasts and continuously updated predictions of ride demand, vehicle supply, and charging needs. These improvements would allow DAC to respond directly to anticipated passenger demand and provide a more adaptive fleet-operating baseline.

In addition, the current case study assumes a homogeneous SAEV fleet and fixed operational rules. Future applications could incorporate variations in battery capacity, charging power, vehicle powertrain, service assignment, and charging-station access to better represent real-world fleet operations and improve the framework's transferability across regions.

The results also reveal an important implementation challenge. Electricity-cost savings and discharging revenues remain small relative to ride-hailing service revenue, while V2G participation may require substantial upfront investment. DAC offers a practical fleet-side strategy without requiring grid-responsive control, whereas V1G provides a feasible near-term extension that captures much of the observed operational, economic, and environmental value using existing unidirectional charging technology. Although Fleet-V2G and Combined-V2G provide broader grid-support potential, their incremental benefits may not yet justify the costs of bidirectional chargers and control systems, utility interconnection, and possible distribution-system upgrades. These costs are not included in the present fleet-profit calculations. Future research should therefore combine the operational framework with infrastructure planning and lifecycle cost analysis to identify the fleet scale, tariff design, and grid-service compensation needed to justify V2G investment.

Overall, this study presents the value of integrating fleet operations with grid-responsive energy management in a dynamic transportation environment and provides insights into the gradual transition from demand-aware fleet coordination to more advanced grid-responsive charging strategies.

## 6. ACKNOWLEDGEMENTS

The authors thank Helena Chandy for her editing (and administrative) support.

## 7. AUTHORS STATEMENT

Lin Su: Writing - original draft, Resources, Investigation, Methodology, Data curation, Conceptualization. Kara M. Kockelman: Writing - review & editing, Supervision, Funding acquisition, Conceptualization.